\documentclass[%
reprint,
superscriptaddress,
nofootinbib,
 amsmath,amssymb,
 aps,
]{revtex4-1}
\usepackage{graphicx}
\usepackage{dcolumn}
\usepackage{bm}
\usepackage[%
bookmarksnumbered,
bookmarksopen,
colorlinks,
citecolor=blue,
linkcolor=blue,
]{hyperref} 

\def\rpi {$\pi^-/\pi^+$~}
\def\es0{$E_{sym}(\rho_0)$~}
\begin{document}


\title{Effects of density-dependent scenarios of in-medium nucleon-nucleon interactions in heavy-ion collisions}

\author{Gao-Feng Wei}\email[ E-mail: ]{wei.gaofeng@gznu.edu.cn}
\affiliation{School of Physics and Electronic Science, Guizhou Normal University, Guiyang 550025, China}
\affiliation{Guizhou Provincial Key Laboratory of Radio Astronomy and Data Processing, Guizhou Normal University, Guiyang 550025, China}
\author{Chang Xu}\email[ E-mail: ]{cxu@nju.edu.cn}
\affiliation{School of Physics, Nanjing University, Nanjing 210008, China}
\author{Wei Xie}
\affiliation{School of Physics and Electronic Science, Guizhou Normal University, Guiyang 550025, China}
\affiliation{Guizhou Provincial Key Laboratory of Radio Astronomy and Data Processing, Guizhou Normal University, Guiyang 550025, China}
\author{Qi-Jun Zhi}
\affiliation{School of Physics and Electronic Science, Guizhou Normal University, Guiyang 550025, China}
\affiliation{Guizhou Provincial Key Laboratory of Radio Astronomy and Data Processing, Guizhou Normal University, Guiyang 550025, China}
\author{Shi-Guo Chen}
\affiliation{School of Physics and Electronic Science, Guizhou Normal University, Guiyang 550025, China}
\author{Zheng-Wen Long}
\affiliation{College of Physics, Guizhou University, Guiyang 550025, China}


\begin{abstract}
	
Using a more reasonable separate density-dependent scenario instead of the total density-dependent scenario for in-medium $nn$, $pp$ and $np$ interactions, we examine effects of differences of in-medium nucleon-nucleon interactions in two density-dependent scenarios on isospin-sensitive observables in central $^{197}$Au+$^{197}$Au collisions at 400 MeV/nucleon. It is shown that the symmetry potentials and resulting symmetry energies in two density-dependent scenarios indeed become to deviate at nonsaturation densities, especially at suprasaturation densities. Naturally, several typical isospin-sensitive observables such as the free neutron-proton ratios and the \rpi ratios in heavy-ion collisions are affected significantly. Moreover, to more physically detect the differences between the nucleon-nucleon interactions in two density-dependent scenarios, we also map the nucleon-nucleon interaction in the separate density-dependent scenario into that in the total density-dependent scenario through fitting the identical constraints for symmetric nuclear matter as well as the identical slope parameter of nuclear symmetry energy at the saturation density. It is shown that two density-dependent scenarios also lead to essentially different 
symmetry potentials especially at high densities although they can lead to the identical equation of state for the symmetry nuclear matter as well as the identical symmetry energy for the isospin asymmetric nuclear matter. Consequently, these isospin-sensitive observables are also appreciably affected by the different density-dependent scenarios of in-medium nucleon-nucleon interactions. Therefore, according to these findings, it is suggested that effects of the separate density-dependent scenario of in-medium nucleon-nucleon interactions should be taken into account when probing the high-density symmetry energy using these isospin-sensitive observables in heavy-ion collisions.

\end{abstract}

\maketitle


\section{introduction}\label{introduction}

Simulations of heavy-ion collisions (HICs) as well as comparisons with the corresponding experiments provide an important tool to explore the properties of strong interacting nucleonic matter at extreme conditions. As the important inputs in simulations of HICs, the density-dependent nucleon-nucleon interactions as well as the resulting nuclear mean field have been paid
much attention in the past few decades~\cite{Pand72,McNe83,Wiri88,Kuts94,UIry97,Das03,Stone03,Dalen04,Zuo05,Chen05,Frit05,LiZH06}. However, the nuclear mean field especially its isovector part, i.e., the symmetry potential, is still incompletely understood at present. Essentially, the symmetry potential is determined by the competition between the isospin singlet and isospin triplet channels of nucleon-nucleon interactions~\cite{Diep03,Zuo91,Xu10a}. The symmetry potential is also found to be sensitive to the in-medium effects such as the in-medium nuclear effective many-body force and the tensor force due to the in-medium $\rho$-meson exchange~\cite{Brown90}. In nonrelativistic models, the in-medium many-body forces effects are usually taken into account by a density-dependent term in the two-body effective interactions~\cite{Vaut72,Onis78,Dech80}, and the relativistic models generate the similar density-dependent term in the two-body effective interactions as in the nonrelativistic models through dressing of the in-medium spinors~\cite{Broc90}. Nevertheless, how exactly does the resulting two-body effective interaction depend on the in-medium nucleon densities remains an open question. For example, a total density-dependent scenario without distinguishing the density dependence for in-medium $nn$, $pp$ and $np$ interactions is usually assumed in the Skyrme, M3Y and Gogny forces and then adopted in some theoretical simulations of HICs~\cite{Kohl76,Dutt86,Tond87,Pear91,Abou92,Fari97,Chab97}. However, within the Brueckner theory~\cite{Brue64,Dabr73,Dabr74}, Brueckner and Dabrowski pointed out that the G-matrix of nucleon-nucleon interactions depends strongly on the respective Fermi momenta of neutrons and protons in isospin asymmetric nuclear matter. 
Actually, the separate density-dependent scenario for in-medium nucleon-nucleon interactions has been used in studying the structure of finite nuclei as well as properties of infinite nuclear matter by some authors such as Negele~\cite{Nege70}, Sprung and Banerjee~\cite{spru71} as well as Brueckner and Dabrowski~\cite{Brue64,Dabr73,Dabr74}.  Of particular interest, authors in Refs.~\cite{Xu10b} and~\cite{Chen14b} employing the Gogny effective interactions, respectively, studied effects of the separate density dependence of in-medium nucleon-nucleon interactions on the symmetry potential and energy, they found consistently that the resulting symmetry potential and energy at nonsaturation densities in the separate density-dependent scenario indeed become to deviate significantly from those in the total density-dependent scenario. Stimulated by these studies, we examine effects of differences of in-medium nucleon-nucleon interactions in these two density-dependent scenarios on isospin-sensitive observables in HICs at intermediate energies. The main purpose of this article is to answer whether one needs to consider the separate density-dependent scenario for in-medium nucleon-nucleon interactions in HICs, especially for probing the density-dependent nuclear symmetry energy using these isospin-sensitive observables in HICs, which is seldom considered in simulations of HICs to our best knowledge.

\section{The Model}\label{Model}
For completeness, we first recall the total density-dependent scenario for in-medium nucleon-nucleon interactions according to the original Gogny effective interaction~\cite{Gogny80},
\begin{eqnarray}\label{Gongy}
v(r)&=&\sum_{i=1,2}(W+BP_{\sigma}-HP_{\tau}-MP_{\sigma}P_{\tau})_{i}e^{-r^{2}/\mu_{i}^{2}}\notag \\
&+&t_{0}(1+x_{0}P_{\sigma}){\big[}\rho{\big(}\frac{{\small\textbf{r}_{i}}+{\small\textbf{r}_{j}}}{{\small{2}}}{\big)}{\big]}^{\alpha}\delta(\textbf{r}_{ij}),
\end{eqnarray}
where $W$, $B$, $H$, $M$, and $\mu$ are five parameters, and $P_{\tau}$ and $P_{\sigma}$ are the isospin and spin exchange operators, respectively; while $\alpha$ is the density-dependent parameter used to mimic in-medium effects of the many-body interactions, particularly, the case with $\alpha=1$ corresponds to an effective density-dependent two-body interaction deduced from a three-body contact interaction in spin-saturated nuclear matter~\cite{Vaut72,Chen14b}. Based on the Hatree-Fock approximation using the original Gogny effective interaction, i.e., Eq.~(\ref{Gongy}), Das $\it{et~al.}$ derived a momentum-dependent interaction (MDI) single-nucleon potential for the Boltzmann-Uehling-Uhlenbeck (BUU) transport model expressed as~\cite{Das03,IBUU}
\begin{eqnarray}
U(\rho,\delta ,\vec{p},\tau ) &=&A_{u}(x)\frac{\rho _{-\tau }}{\rho _{0}}%
+A_{l}(x)\frac{\rho _{\tau }}{\rho _{0}}  \notag \\
&+&B(\frac{\rho }{\rho _{0}})^{\sigma }(1-x\delta ^{2})-8\tau x\frac{B}{%
	\sigma +1}\frac{\rho ^{\sigma -1}}{\rho _{0}^{\sigma }}\delta \rho
_{-\tau }
\notag \\
&+&\frac{2C_{l}}{\rho _{0}}\int d^{3}p^{\prime }\frac{f_{\tau }(%
	\vec{p}^{\prime })}{1+(\vec{p}-\vec{p}^{\prime })^{2}/\Lambda ^{2}}
\notag \\
&+&\frac{2C_{u}}{\rho _{0}}\int d^{3}p^{\prime }\frac{f_{-\tau }(%
	\vec{p}^{\prime })}{1+(\vec{p}-\vec{p}^{\prime })^{2}/\Lambda ^{2}},
\label{MDIU}
\end{eqnarray}%
where $\tau=1/2$ for neutrons and $-1/2$ for protons; while parameters $A_{u}(x)$ and $A_{l}(x)$ are determined as
\begin{equation}\label{MDIp}
A_{u}(x)=-95.98 - x\frac{2B}{\sigma+1},~~A_{l}(x)=-120.57 + x\frac{2B}{\sigma+1}.
\end{equation}
Here, the parameter $x$ is related to the spin(isospin)-dependent parameter $x_{0}$ via $x=(1+2x_{0})/3$ in the density-dependent term of original Gogny effective interactions, which controls the relative contributions of the density-dependent term to the total energy in the isospin singlet channel $[\propto(1+x_{0})\rho^{\alpha+1}]$ and triplet channel $[\propto(1-x_{0})\rho^{\alpha+1}]$~\cite{Gogny80}. Therefore, varying the $x$ parameter can cover uncertainties of the spin(isospin) dependence of in-medium many-body forces which are responsible for the divergent density dependence of nuclear symmetry energy in the Gogny Hatree-Fock calculations~\cite{Stone03,Xu10a,Chen14b}. However, it should be emphasized that the $x$ parameter does not affect the equation of state of symmetric nuclear matter as well as the symmetry energy at the saturation density due to the contributions of different channels are cancelled out exactly, i.e., $\propto(1+x_{0})\rho^{\alpha+1}+(1-x_{0})\rho^{\alpha+1}=2\rho^{\alpha+1}$. The parameters $B=106.35$~MeV and $\sigma=4/3$ in the MDI single-nucleon potential are related to $t_{0}$ and $\alpha$ in the original Gogny effective interactions via $t_{0}=\frac{8}{3}\frac{B}{\sigma+1}\frac{1}{\rho_{0}^{\sigma}}$ and $\sigma=\alpha+1$~\cite{Xu10a,Xu10b}. While $C_{u}=-103.4$~MeV and $C_{l}=-11.7$~MeV are the interaction strength parameters for a nucleon with isospin $\tau$ interacting, respectively, with unlike and like nucleons in the nuclear matter, and thus account for the momentum dependence of the single-nucleon potential. These parameters are all obtained by fitting the reached consensuses on properties of nuclear matter at the saturation density $\rho_{0}=0.16$~fm$^{-3}$ including the binding energy $E_{0}(\rho_{0})=-16$~MeV, the incompressibility $K_{0}=212$~MeV for symmetric nuclear matter, as well as the symmetry energy $E_{sym}(\rho_{0})=30.5$~MeV, for more details about the MDI interaction, see, e.g., Refs.~\cite{Das03,IBUU}.

While for the separate density-dependent scenario for in-medium nucleon-nucleon interactions, we follow the Ref.~\cite{Xu10b} to replace the density-dependent term in the original Gongy effective interaction by the following density-dependent term
\begin{equation}\label{den-term}
V_{D}=t_{0}(1+x_{0}P_{\sigma})[\rho_{\tau_{i}}(\textbf{r}_{i}){\small +}\rho_{\tau_{j}}(\textbf{r}_{j})]^{\alpha}\delta(\textbf{r}_{ij}).
\end{equation}
Here, the interaction explicitly depends on densities of two nucleons at positions $\textbf{r}_{i}$ and $\textbf{r}_{j}$ instead of the total density of two-nucleon central position $(\textbf{r}_{i}+\textbf{r}_{j})/2$ as in the original Gogny effective interaction. With this density-dependent scenario for in-medium nucleon-nucleon interactions, the resulting single-nucleon potential labelled as the improved MDI single-nucleon potential (IMDI) is changed as~\cite{Xu10b}
\begin{eqnarray}
U^{'}(\rho,\delta ,\vec{p},\tau ) &=&A^{'}_{u}(x)\frac{\rho _{-\tau }}{\rho _{0}}%
+A^{'}_{l}(x)\frac{\rho _{\tau }}{\rho _{0}}+\frac{B}{2}{\big(}\frac{2\rho_{\tau} }{\rho _{0}}{\big)}^{\sigma }(1-x)  \notag \\
&+&\frac{2B}{%
	\sigma +1}{\big(}\frac{\rho}{\rho _{0}}{\big)}^{\sigma }(1+x)\frac{\rho_{-\tau}}{\rho}{\big[}1+(\sigma-1)\frac{\rho_{\tau}}{\rho}{\big]}
\notag \\
&+&\frac{2C_{l }}{\rho _{0}}\int d^{3}p^{\prime }\frac{f_{\tau }(%
	\vec{p}^{\prime })}{1+(\vec{p}-\vec{p}^{\prime })^{2}/\Lambda ^{2}}
\notag \\
&+&\frac{2C_{u }}{\rho _{0}}\int d^{3}p^{\prime }\frac{f_{-\tau }(%
	\vec{p}^{\prime })}{1+(\vec{p}-\vec{p}^{\prime })^{2}/\Lambda ^{2}},
\label{IMDIU}
\end{eqnarray}%
and the corresponding parameters $A_{u}(x)$ and $A_{l}(x)$ are changed as
\begin{eqnarray}
A^{'}_{u}(x)&=&-95.98 - \frac{2B}{\sigma+1}\big{[}1-2^{\sigma-1}(1-x)\big{]},  \\
A^{'}_{l}(x)&=&-120.57 + \frac{2B}{\sigma+1}\big{[}1-2^{\sigma-1}(1-x)\big{]}.
\end{eqnarray}
It should be mentioned that the properties of symmetric nuclear matter are not changed from the MDI interaction to the IMDI interaction due to the isospin scalar potentials $U_{0}(\rho,0,\vec{p},\tau)=U^{'}_{0}(\rho,0,\vec{p},\tau)$ by setting $\delta=0$ and $\rho_{n}=\rho_{p}=\frac{1}{2}\rho$. While for the isospin asymmetric nuclear matter, the properties are expected to change from the MDI interaction to the IMDI interaction. 

\begin{figure}[t]
	\includegraphics[width=0.8\columnwidth]{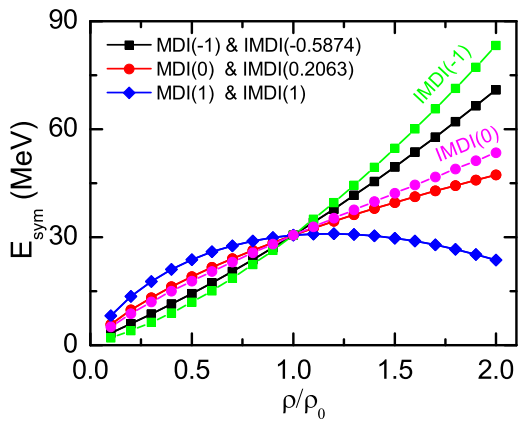}
	\caption{(Color online) The density dependencies of nuclear symmetry energies calculated from the MDI and IMDI single-nucleon potentials.} 
	\label{symene}
\end{figure}

Shown in Fig.~\ref{symene} are the density dependencies of nuclear symmetry energy calculated from the MDI and IMDI single-nucleon potentials. For parameter $x=1$, it is seen that the symmetry energy calculated from the IMDI single-nucleon potential is the same as that calculated from the MDI single-nucleon potential. This is because the fourth term in the IMDI single-nucleon potential is zero with $x=1$ while other terms are unchanged as in the MDI single-nucleon potential. However, for parameters of $x=-1$ and $0$, the symmetry energies calculated from the IMDI single-nucleon potential become stiffer (softer) at suprasaturation (subsaturation) densities compared to those calculated from the MDI single-nucleon potential. Undoubtedly, these differences of the symmetry energy are resulting from different symmetry potentials generated in the MDI and IMDI single-nucleon potentials. 
It should be emphasized that even with the same symmetry energies, the corresponding symmetry potentials could be very different due to the fact that the symmetry potentials depend not only on the nucleon density but also on the nucleon momentum or energy. 
Therefore, to physically distinguish the symmetry potentials derived from the MDI and IMDI single-nucleon potentials, it is useful to map the nucleon-nucleon interaction in the separate density-dependent scenario (i.e., IMDI single-nucleon potential) into that in the total density-dependent scenario (i.e., MDI single-nucleon potential). This is carried out by fitting the identical constraints for symmetric nuclear matter as well as the identical slope parameter of symmetry energy at $\rho_{0}$, the corresponding results are also shown in Fig.~\ref{symene}. It is seen that the symmetry energy calculated from the IMDI single-nucleon potential with mapped parameters $x=-0.5874$ and $0.2063$, respectively, is completely identical with that calculated from the MDI single-nucleon potential with parameters $x=-1$ and $0$. 

\begin{figure*}[htb]
	\includegraphics[width=0.8\textwidth]{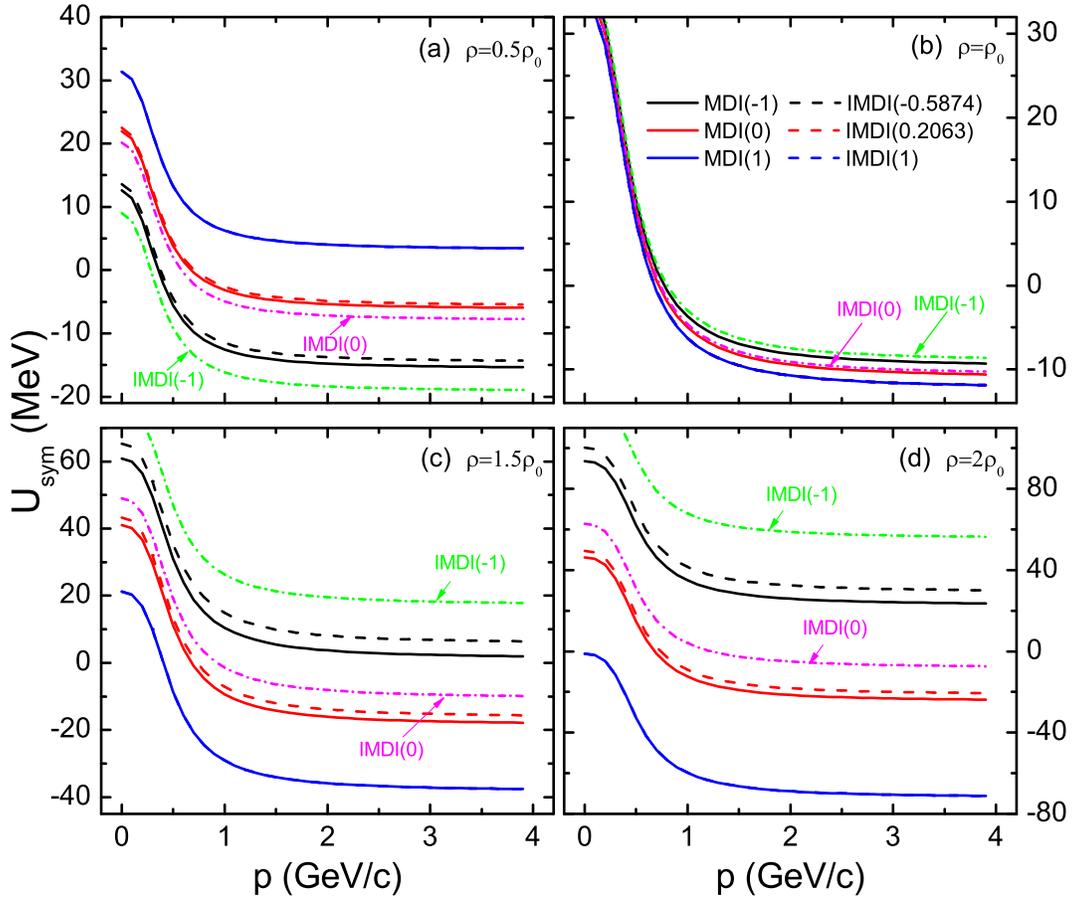}
	\caption{(Color online) The momentum dependencies of symmetry potentials calculated from the MDI and IMDI single-nucleon potentials.} \label{sympot}
\end{figure*}

Shown in Fig.~\ref{sympot} are the momentum-dependent symmetry potentials at $\rho=0.5\rho_{0}$, $\rho_{0}$, $1.5\rho_{0}$ and $2\rho_{0}$ calculated from the MDI and IMDI single-nucleon potentials with parameters $x=-1$, $0$ and $1$ as well as the mapped parameters $x=-0.5874$ and $0.2063$ used in the IMDI single-nucleon potential. Again, with parameter $x=1$, the symmetry potentials are completely identical to each other in calculations using the MDI and IMDI single-nucleon potentials at either low densities or high densities. While for parameter $x=-1$, the symmetry potentials are significantly stronger especially at high densities in calculations using the IMDI single-nucleon potential compared to those in calculations using the MDI single-nucleon potential. However, for the case of parameter $x=0$, the symmetry potentials calculated from the IMDI single-nucleon potential change from strong to weak (weak to strong) at suprasaturation (subsaturation) densities with the increase of nucleons momenta, compared to those calculated from the MDI single-nucleon potential.
As to the mapped symmetry potentials calculated from the IMDI single-nucleon potential with parameters $x=-0.5874$ and $0.2063$, it can be observed that they are also appreciably different from those in calculations using the MDI single-nucleon potetnial with parameters $x=-1$ and $0$, respectively. Actually, according to the relation between the symmetry energy and the single-nucleon potential~\cite{Xu10a,Brue64,Dabr73,Dabr74}, i.e.,
\begin{equation}\label{rela}
E_{sym}(\rho)\approx\frac{1}{3}t(k_{F})+\frac{1}{6}\frac{\partial{U_{0}}}{\partial{k}}\mid_{k_{F}}k_{F}+\frac{1}{2}U_{sym}(k_{F}),
\end{equation}
with $t(k)$ denotes the nucleon kinetic energy and $k_{F}$ represents the Fermi momentum of nucleons in symmetry nuclear matter, one can know that these are mainly due to the differences between the symmetry potential calculated from the MDI single-nucleon potential and the mapped symmetry potential calculated from the IMDI single-nucleon potential. 
Therefore, effects of two density-dependent scenarios for in-medium nucleon-nucleon interactions on isospin-sensitive observables in HICs can be reflected through examining effects of the symmetry potential calculated from the MDI single-nucleon potential and the mapped symmetry potential calculated from the IMDI single-nucleon potential on these observables in HICs. Nevertheless, it should be emphasized that the differences between the MDI symmetry potential and the IMDI mapped symmetry potential are essentially resulting from different density-dependent scenarios because the momentum-dependent but the $x$ parameter independent $C$ terms in Eq.~(\ref{IMDIU})  are completely identical to that in Eq.~(\ref{MDIU}). As a result, according to the formula of nucleon effective mass, i.e., 
\begin{equation}\label{effective-mass}
m^{*}_{\tau}/m=\big{[}1+\frac{m}{k_{\tau}}\frac{dU_{\tau}}{dk}\big{]}^{-1},
\end{equation}
which is only related to the $C$ terms in Eq.~(\ref{MDIU}) and Eq.~(\ref{IMDIU}), one can know that the nucleon effective mass as well as its isospin splitting are not changed from the MDI interaction to the IMDI interaction.

\section{Results and Discussions}\label{Results and Discussions}

Now, we compare effects of the symmetry potential calculated from the MDI single-nucleon potential and the mapped symmetry potential calculated from the IMDI single-nucleon potential on isospin-sensitive observables in HICs. As comparisons, we also include the corresponding results calculated from the IMDI single-nucleon potential with parameters $x=-1$ and $0$ in the following discussions. 

\begin{figure}[t]
	\includegraphics[width=0.8\columnwidth]{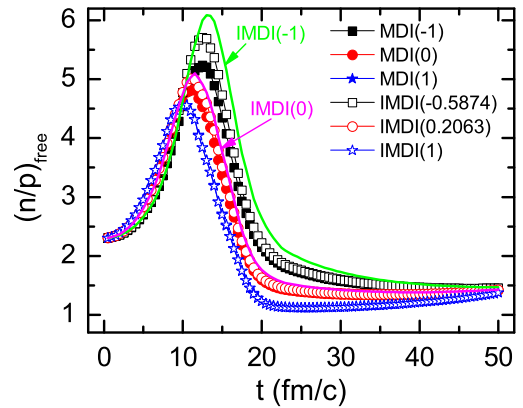}
	\caption{(Color online) Evolutions of free neutron-proton ratios in calculations with the MDI and IMDI single-nucleon potentials.} \label{nprt}
\end{figure}

\begin{figure}[t]
	\includegraphics[width=0.8\columnwidth]{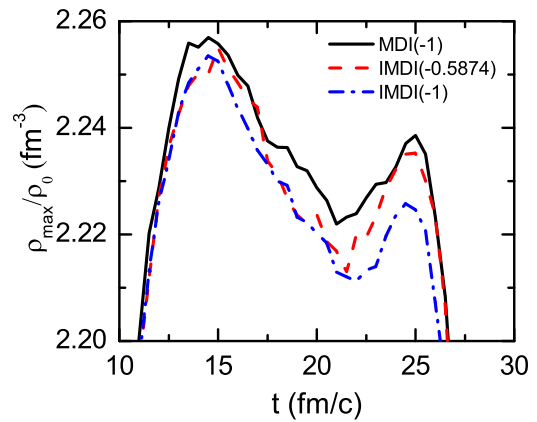}
	\caption{(Color online) Evolutions of reduced maximum densities $\rho_{{\rm max}}/\rho_{0}$ reached at the maximum compression stages in collisions with the MDI and IMDI single-nucleon potentials.} \label{aveDen}
\end{figure}

\begin{figure}[htb]
	\includegraphics[width=0.8\columnwidth]{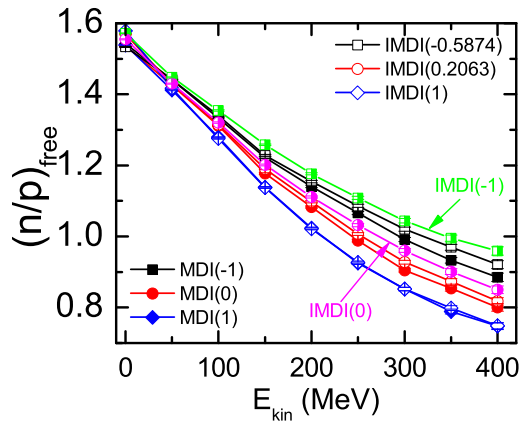}
	\caption{(Color online) Kinetic energy distributions of free neutron-proton ratios at the end of reactions with the MDI and IMDI single-nucleon potentials.} \label{Ekindis}
\end{figure}

Show in Fig.~\ref{nprt} are the free neutron-proton ratios generated in central $^{197}$Au + $^{197}$Au collisions at 400 MeV/nucleon where the free neutrons and protons are defined as those with local densities less than $\rho_{0}/8$. First, as expected, with parameter $x=1$, the free neutron-proton ratio generated in simulations using the IMDI single-nucleon potential is completely identical with that in simulations using the MDI single-nucleon potential. Second, with parameter $x=-1$ and $0$, the free neutron-proton ratios generated in simulations using the IMDI single-nucleon potential are larger especially at the compression stage compared to those in simulations using the MDI single-nucleon potential, reflecting the free neutron-proton ratios are indeed sensitive to high-density behaviors of nuclear symmetry potential and/or energy because the stronger positive symmetry potentials at high densities get more neutrons to be spread but more protons to be gathered. Certainly, for the case of parameter $x=0$, the competition of symmetry potentials at high densities between low nucleon momentum and high nucleon momentum as aforementioned causes the observed effects to be not so obvious as those in the case of parameter $x=-1$. However, it should be emphasized that, besides the different density-dependent scenarios of in-medium nucleon-nucleon interactions, these effects are also resulting from the different symmetry energy settings because the slope values $L$ of nuclear symmetry energy at $\rho_{0}$ are completely different although the identical parameters $x$ are used in the MDI and IMDI single-nucleon potentials. Therefore, to more physically detect the effects of differences of in-medium nucleon-nucleon interactions in two density-dependent scenarios on the free neutron-proton ratios, we compare the free neutron-proton ratios generated in calculations using the MDI single-nucleon potential with parameters $x=-1$ and $0$ and those in calculation using the IMDI single-nucleon potential with mapped parameters $x=-0.5874$ and $0.2063$ as well as those in calculations using the IMDI single-nucleon potential with parameters $x=-1$ and $0$. As shown also in Fig.~\ref{nprt}, the free neutron-proton ratios in calculations using the IMDI single-nucleon potentials with mapped parameters $x=-0.5874$ and $0.2063$ are obviously lower than those in calculations using the IMDI single-nucleon potentials with parameters $x=-1$ and $0$, respectively. Certainly, the values of free neutron-proton ratios in calculations using the IMDI single-nucleon potentials with mapped parameters $x=-0.5874$ and $0.2063$ are also larger than those in calculations using the MDI single-nucleon  potentials with parameters $x=-1$ and $0$, respectively. Actually, the symmetry energy is not directly entering into the reaction process, and thus does not directly affect the free neutron-proton ratios. On the contrary, the different symmetry energy is originated from different symmetry potential, i.e., the isovector part of single-nucleon potential under different $x$ parameter settings. Naturally, these different single-nucleon potentials dominate the different reaction dynamics as well as different free neutron-proton ratios which thus indirectly reflect different symmetry energy settings. This can be confirmed by comparing the reduced maximum densities $\rho_{{\rm max}}/\rho_{0}$ reached at the maximum compression stages in collisions with the IMDI single-nucleon potential under setting two different parameters $x=-1$ (IMDI(-1)) and $x=-0.5874$ (IMDI(-0.5874)) as shown in Fig.~\ref{aveDen}. Nevertheless, for the case with the MDI single-nucleon potential under setting $x=-1$ (MDI(-1)) and that with the IMDI single-nucleon potential under setting $x=-0.5874$ (IMDI(-0.5874)), the resulting $\rho_{{\rm max}}/\rho_{0}$ are also different although their corresponding symmetry energies are completely identical as shown in Fig.~\ref{symene}. Actually, according to the formula (\ref{rela}) as well as the Fig.~\ref{sympot}, this is exactly the difference of in-medium nucleon-nucleon interactions in two density-dependent scenarios that leads to the different $\rho_{{\rm max}}/\rho_{0}$ in collisions, and thus generates different free neutron-proton ratios. Correspondingly, the kinetic energy distributions of free neutron-proton ratios at the end of reactions are also affected by the density-dependent scenarios of in-medium nucleon-nucleon interactions as shown in Fig.~\ref{Ekindis}. Therefore, effects of the separate density-dependent scenario of in-medium nucleon-nucleon interactions should be carefully considered in studies of using the free neutron-proton ratio as a probe of nuclear symmetry energy especially at high densities.

\begin{figure}[t]
	\includegraphics[width=0.8\columnwidth]{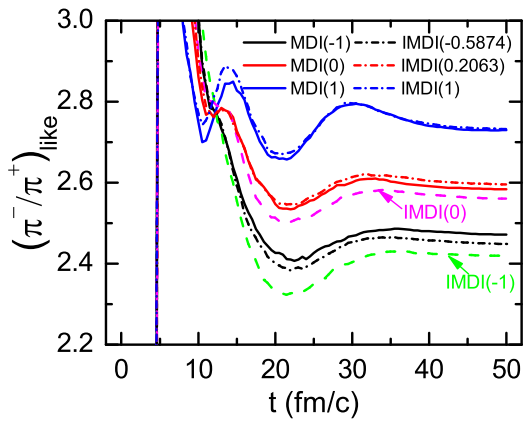}
	\caption{(Color online) Evolutions of \rpi ratios in calculations with the MDI and IMDI single-nucleon potentials.} \label{piratio}
\end{figure}

On the other hand, according to the production mechanism of pions, i.e., pions are produced mainly at the compression stages during collisions and $\pi^{-}$ is mainly from $nn$ inelastic collisions but $\pi^{+}$ mainly from $pp$ inelastic collisions, one naturally expects that the effects of density-dependent scenarios of in-medium nucleon-nucleon interactions also hold for the \rpi ratio, which has been indicated to be very sensitive to the symmetry energy and potential at high densities~\cite{Uma98,BALi02,Ditoro05,FOPI,BALi08} but still affected by some incompletely known uncertainties~\cite{Hong14,Song15,BALi15a,BALi15b,Hen15,Cozma16,trans1,trans2,trans3}. In HICs at intermediate energies, pions are produced during collisions mostly from the decay of $\Delta(1232)$, therefore, it is useful to examine the effects of density-dependent scenarios of in-medium nucleon-nucleon interactions on dynamic pion ratio $(\pi^{-}/\pi^{+})_{{\rm like}}$, i.e.,
\begin{equation}\label{pilike}
(\pi^{-}/\pi^+{})_{{\rm like}}=\frac{\pi^{-}+\Delta^{-}+\frac{1}{3}\Delta^{0}}{\pi^{+}+\Delta^{++}+\frac{1}{3}\Delta^{+}}.
\end{equation}
Certainly, because all the $\Delta$ resonances will eventually decay into nucleons and pions, the ratio $(\pi^{-}/\pi^{+})_{{\rm like}}$ will naturally become to the free \rpi ratio at the end of reactions. Shown in Fig.~\ref{piratio} are the evolutions of $(\pi^{-}/\pi^{+})_{{\rm like}}$ ratios generated in central $^{197}$Au + $^{197}$Au collisions at 400 MeV/nucleon. First, it is obvious to see that the \rpi ratio is indeed sensitive to the density dependence of nuclear symmetry energy regardless of using the IMDI single-nucleon potential or the MDI single-nucleon potential, and a softer symmetry energy usually leads to a higher \rpi ratio, reflecting a more neutron-rich participant region formed in the reaction~\cite{Uma98,BALi02,Ditoro05,FOPI,BALi08}. Second, with parameter $x=1$, the \rpi ratio calculated from the IMDI single-nucleon potential is completely identical with that calculated from the MDI single-nucleon potential. Third, for the case of identical parameter $x=-1$, it is seen that the stronger positive symmetry potential at high densities gets the \rpi ratio in calculations using the IMDI single-nucleon potential to be significantly smaller than that in calculations using the MDI single-nucleon potential. Actually, due to the stronger positive symmetry potential with parameter $x=-1$ causes more neutrons to be spread but more protons to be gathered, naturally, according to the production mechanism of pions, i.e., $\pi^{-}$ is produced mainly from the channel $n+n\rightarrow\pi^{-}+p$ but $\pi^{+}$ from the channel $p+p\rightarrow\pi^{+}+p$, we can observe a smaller \rpi ratios in calculations using the IMDI single-nucleon potential. Certainly, due to the symmetry potential calculated from the IMDI single-nucleon potential with parameter $x=0$ changes from strong to weak at suprasaturation densities with the increase of nucleons momenta compared to that calculated from the MDI single-nucleon potential with parameter $x=0$, we can also see that the differences of \rpi ratios in this case are not as larger as those in the case of parameter $x=-1$. Again, these effects are resulting from both different density-dependent scenarios and different symmetry energy settings. While comparing the \rpi ratios in calculations using the IMDI single-nucleon potential with mapped parameters $x=-0.5874$ and $0.2063$ with those in calculations using the MDI single-nucleon potential with parameters $x=-1$ and $0$, respectively, we find again the similar observations as those in free neutron-proton ratios due to the IMDI mapped symmetry potential is after all appreciably different from the MDI symmetry potential. Especially, due to the differences of the IMDI mapped symmetry potential with parameter $x=-0.5874$ and the MDI symmetry potential with parameter $x=-1$ are relative large at both low nucleon momentum and high nucleon momentum, the corresponding effects on the \rpi ratios are relative appreciable. Certainly, it should be emphasized that this effect is independent of nuclear symmetry energy but is exactly resulting from different density-dependent scenarios of in-medium nucleon-nucleon interactions. Therefore, effects of the separate density-dependent scenario of in-medium nucleon-nucleon interactions should also be carefully considered in studies of using the \rpi ratio as a probe of nuclear symmetry energy especially at high densities. 

Before ending this part, we give two useful remarks. 

First, although there are currently no physical studies based on first principles to illustrate more accuracy of the separate density-dependent scenario, some results relevant to nuclear structure studies have been shown to yield very satisfactory agreement with the corresponding experiments such as the binding energies, single-particle energies, and electron scattering cross sections for $^{16}$O, $^{40}$Ca, $^{48}$Ca, $^{90}$Zr, and $^{208}$Pr~\cite{Nege70,Dugues03}. Moreover, as indicated in Ref.~\cite{Dugues03}, the separate density dependence of effective two-body interactions is originated from the renormalization of multibody force effects, and the latter may extend the density dependence of effective interactions for calculations beyond the mean-field approximation and open a new freedom in the effective interactions.

Second, it is well known that the double neutron-proton and/or \rpi ratios from two reaction systems have the advantage of reducing both systematic errors and the influences of isoscalar potentials in HICs~\cite{LiBA06,Wei19}. This could enlarge the contribution of the isovector potentials and better discriminate between the two scenarios. Therefore, the double ratios of these observables from two reactions as well as the cross examinations of these observables using various experimental data such as the FOPI data~\cite{FOPI} and that from the symmetry energy measurement experiment at RIBF-RIKEN in Japan~\cite{Shane15} could be good candidates in probing the effects of density-dependent scenarios in HICs in future.

\section{Summary}\label{Summary}
In conclusion, we have studied effects of differences of in-medium nucleon-nucleon interactions in the separate and total density-dependent scenarios on isospin-sensitive observables in HICs within a transport model. Consistent with the previous studies, the nuclear symmetry energy and potential at nonsaturation densities in the separate density-dependent scenario indeed become to deviate significantly from those in the total density-dependent scenario for the identical $x$ parameters except for the parameter $x=1$. Two typical isospin-sensitive observables including the free neutron-proton ratios and the \rpi ratios in HICs are affected significantly. Nevertheless, it should be emphasized that these effects are resulting from both the different symmetry energy settings and the different density-dependent scenarios of in-medium nucleon-nucleon interactions. Therefore, to more physically detect the differences of in-medium nucleon-nucleon interactions as well as the resulting symmetry potential in two density-dependent scenarios, we have also mapped the nucleon-nucleon interaction in the separate density-dependent scenario into that in the total density-dependent scenario through fitting the identical constraints for symmetric nuclear matter as well as the identical slope parameter of symmetry energy at the saturation density. It is shown that the mapped symmetry potentials calculated from the IMDI single-nucleon potential indeed deviate from those in calculations using the MDI single-nucleon potential especially at high densities. Consequently, these isospin-sensitive observables in HICs could also be appreciably affected. Therefore, according to these findings as well as the Brueckner theory and previous findings in nuclear structure studies, we conclude that effects of the separate density-dependent scenario for in-medium nucleon-nucleon interactions might be very important and thus should be taken into account when probing the high-density symmetry energy using these isospin-sensitive observables in HICs.

\begin{acknowledgments}
This work is supported by the National Natural Science Foundation of China under grant Nos.11965008, 11822503,  U1731218, 11847102, and Guizhou Provincial Science and Technology Foundation under Grant No.[2020]1Y034, and the PhD-funded project of Guizhou Normal university (Grant No.GZNUD[2018]11).
\end{acknowledgments}


\begin{thebibliography}{99}
	
\bibitem{Pand72}V. R. Pandharipande, V. K. Garde, Phys. Lett. B \textbf{39}, 608 (1972).

\bibitem{McNe83}J. A. McNeil, J. R. Shepard, and S. J. Wallace, Phys. Rev. Lett. \textbf{50}, 1439 (1983).

\bibitem{Wiri88}R. B. Wiringa, V. Fiks, and A. Fabrocini, Phys. Rev. C \textbf{38}, 1010 (1988).

\bibitem{Kuts94}M. Kutschera, Phys. Lett. B \textbf{340}, 1 (1994).

\bibitem{UIry97}S. Ulrych and H. M\"{u}ther, Phys. Rev. C \textbf{56}, 1788 (1997).

\bibitem{Das03} C. B. Das, S. Das Gupta, C. Gale, and B. A. Li, Phys. Rev. C \textbf{67}, 034611 (2003).

\bibitem{Stone03}J. R. Stone, J. C. Miller, R. Koncewicz, P. D. Stevenson, and M. R. Strayer, Phys. Rev. C \textbf{68}, 034324 (2003).

\bibitem{Dalen04}E. N. E. van Dalen, C. Fuchs, and A. Faessler, Nucl. Phys. A \textbf{744}, 227 (2004).

\bibitem{Zuo05}W. Zuo, L. G. Cao, B. A. Li, U. Lombardo, and C. W. Shen, Phys. Rev. C \textbf{72}, 014005 (2005).

\bibitem{Chen05}L. W. Chen, C. M. Ko, and B. A. Li, Phys. Rev. C \textbf{72}, 064606 (2005).

\bibitem{Frit05}S. Fritsch, N. Kaiser, and W. Weise, Nucl. Phys. A \textbf{750}, 259 (2005).

\bibitem{LiZH06}Z. H. Li, L. W. Chen, C. M. Ko, B. A. Li, and H. R. Ma, Phys. Rev. C \textbf{74}, 044613 (2006).

\bibitem{Diep03}A. E. L. Dieperink, Y. Dewulf, D. Van Neck, M. Waroquier, and V. Rodin, Phys. Rev. C \textbf{68}, 064307 (2003).

\bibitem{Zuo91}W. Zuo, L. G. Cao, B. A. Li, U. Lombardo, and C. W. Shen, Phys. Rev. C \textbf{72}, 014005 (2005).

\bibitem{Xu10a}C. Xu, B. A. Li, Phys. Rev. C \textbf{81}, 064612 (2010).

\bibitem{Brown90}G. E. Brown, M. Rho, Phys. Lett. \textbf{237}, 3 (1990), Phys. Rev. Lett. \textbf{66}, 2720 (1991), Phys. Rep. \textbf{396}, 1 (2004).

\bibitem{Vaut72}D. Vautherin, D. M. Brink, Phys. Rev. C \textbf{5}, 626 (1972).

\bibitem{Onis78}N. Onishi, J. W. Negele, Nucl. Phys. A \textbf{301}, 336 (1978).

\bibitem{Dech80}J. Decharg\'{e}, D. Gogny, Phys. Rev. C \textbf{21}, 1568 (1980).

\bibitem{Broc90}R. Brockmann, R. Machleidt, Phys. Rev. C \textbf{42}, 1965 (1990).

\bibitem{Kohl76}S. K\"{o}hler, Nucl. Phys. A \textbf{258}, 301 (1976).

\bibitem{Dutt86}A. K. Dutta, J. P. Arcoragi, J. M. Pearson, R. Behrman, and F. Tondeur, Nucl. Phys. A \textbf{458}, 77 (1986).

\bibitem{Tond87}F. Tondeur, A. K. Dutta, J. M. Pearson, and R. Behrman, Nucl. Phys. A \textbf{470}, 93 (1987).

\bibitem{Pear91}J. M. Pearson, Y. Aboussir, A. K. Dutta, R. C. Nayak, M. Farine, and F. Tondeur, Nucl. Phys. A \textbf{528}, 1 (1991).

\bibitem{Abou92}Y. Aboussir, J. M. Pearson, A. K. Dutta, and F. Tondeur, Nucl. Phys. A \textbf{549}, 155 (1992).

\bibitem{Fari97}M. Farine, J. M. Pearson, and F. Tondeur, Nucl. Phys. A \textbf{615}, 135 (1997).

\bibitem{Chab97}E. Chabanat, P. Bonche, P. Haensel, J. Meyer, and R. Schaeffer, Nucl. Phys. A \textbf{627}, 710 (1997).

\bibitem{Brue64}K. A. Brueckner, J. Dabrowski, Phys. Rev. \textbf{134}, B722 (1964).

\bibitem{Dabr73}J. Dabrowski, P. Haensel, Phys. Rev. C \textbf{7}, 916 (1973).

\bibitem{Dabr74}J. Dabrowski, P. Haensel, Can. J. Phys. \textbf{52}, 1768 (1974).

\bibitem{Nege70}J. W. Negele, Phys. Rev. C \textbf{1}, 1260 (1970).

\bibitem{spru71}D. W. L. Sprung, P. K. Banerjee, Nucl. Phys. A \textbf{168}, 273 (1971).

\bibitem{Xu10b}C. Xu, B. A. Li, Phys. Rev. C \textbf{81}, 044603 (2010).

\bibitem{Chen14b}L. W. Chen, C. M. Ko, B. A. Li, C. Xu, and J. Xu, Eur. Phys. J. A \textbf{50}, 29 (2014).

\bibitem{Gogny80}J. Decharg\'{e}, D. Gogny, Phys. Rev. C \textbf{21}, 1568 (1980).

\bibitem{IBUU}B. A. Li, C. B. Das, S. Das Gupta, and C. Gale, Phys. Rev. C \textbf{69}, 011603(R) (2004).

\bibitem{Uma98}V. S. Uma Maheswari, C. Fuchs, Amand Faessler, Z. S. Wang, and D. S. Kosov, Phys. Rev. C \textbf{57}, 922 (1998).

\bibitem{BALi02}B. A. Li, Phys. Rev. Lett. \textbf{88}, 192701 (2002).

\bibitem{Ditoro05}V. Baran, M. Colonna, V. Greco, and M. Di Toro, Phys. Rep. \textbf{410}, 335 (2005).

\bibitem{FOPI} W. Reisdorf {\it et al.} (FOPI Collaboration), Nucl. Phys. A \textbf{781}, 459 {2007}; Nucl. Phys. A \textbf{848}, 366 (2010); Nucl. Phys. A \textbf{876}, 1 (2012).

\bibitem{BALi08}B. A. Li, L. W. Chen, and C. M. Ko, Phys. Rep. \textbf{464}, 113 (2008).

\bibitem{Hong14}J. Hong, P. Danielewicz, Phys. Rev. C \textbf{90}, 024605 (2014).

\bibitem{Song15}T. Song, C. M. Ko, Phys. Rev. C \textbf{91}, 014901 (2015).

\bibitem{BALi15a}B. A. Li, W. J. Guo, and Z. Z Shi, Phys. Rev. C \textbf{91}, 044601 (2015).

\bibitem{BALi15b}B. A. Li, Phys. Rev. C \textbf{92}, 034603 (2015).

\bibitem{Hen15}O. Hen, B. A. Li, W. J. Guo, L. B. Weinstein, and Eli Piasetzky, Phys. Rev. C \textbf{91}, 025803 (2015).

\bibitem{Cozma16}M. D. Cozma, Phys. Lett. B \textbf{753}, 166 (2016).

\bibitem{trans1}J. Xu, L. W. Chen, M. B. Tsang, H. Wolter, Y. X. Zhang, J. Aichelin, M. Colonna, D. Cozma, P. Danielewicz, Z. Q. Feng, A. L. Fevre, T. Gaitanos, C. Hartnack, K. Kim, Y. Kim, C. M. Ko, B. A. Li, Q. F. Li, Z. X. Li, P. Napolitani, A. Ono, M. Papa, T. Song, J. Su, J. L. Tian, N. Wang, Y. J. Wang, J. Weil, W. J. Xie, F. S. Zhang, and G. Q. Zhang, Phys. Rev. C \textbf{93}, 044609 (2016).

\bibitem{trans2}Y. X. Zhang, Y. J. Wang, M. Colonna, P. Danielewicz, A. Ono, B. Tsang, H. Wolter, J. Xu, L. W. Chen, D. Cozma, Z. Q. Feng, S. D. Gupta, N. lkeno, C. M. Ko, B. A. Li, Q. F. Li, Z. X. Li, S. Mallik, Y. Nara, T. Ogawa, A. Ohnishi, D. Oliinychenko, M. Papa, H. Petersen, J. Su, T. Song, J. Weil, N. Wang, F. S. Zhang, and Z. Zhang, Phys. Rev. C \textbf{97}, 034625 (2018).

\bibitem{trans3}A. Ono, J. Xu, M. Colonna, P. Danielewicz, C. M. Ko, M. B. Tsang, Y. J. Wang, H. Wolter, Y. X. Zhang, L. W. Chen, D. Cozma, H. Elfner, Z. Q. Feng, N. Ikeno, B. A. Li, S. Mallik, Y. Nara, T. Ogawa, A. Ohnishi, D. Oliinychenko, J. Su, T. Song, F. S. Zhang, and Z. Zhang, Phys. Rev. C \textbf{100}, 044617 (2019).

\bibitem{Dugues03} T. Duguet, P. Bonche, Phys. Rev. C \textbf{67}, 054308 (2003).

\bibitem{LiBA06}B. A. Li, L. W. Chen, G. C. Yong, and W. Zuo, Phys. Lett. B \textbf{634}, 378 (2006).

\bibitem{Wei19}G. F. Wei, G. C. Yong, L. Ou, Q. J. Zhi, Z. W. Long, and X. H. Zhou, Phys. Rev. C \textbf{98}, 024618 (2018).

\bibitem{Shane15}R. Shane, A. B. McIntosh, T. Isobe, W. G. Lynch, H. Baba, J. Barney, Z. Chajecki, M. Chartier, J. Estee, M. Famiano, B. Hong, K. Ieki, G. Jhang, R. Lemmon, F. Lua, T. Murakami, N. Nakatsuka, M. Nishimura, R. Olsen, W. Powell, H. Sakurai, A. Taketani, S. Tangwancharoen, M. B. Tsang, T. Usukura, R. Wang, S. J. Yennello, and J. Yurkon, 
et al., S$\pi$RIT: A time-projection chamber for symmetry-energy studies. Nuclear Instruments and Methods in Physics Research A \textbf{784}, 513 (2015). 
	
\end{thebibliography}
\end{document}